\documentclass[twocolumn,showpacs,preprintnumbers,amsmath,amssymb]{revtex4}

\usepackage{graphicx}
\usepackage{dcolumn}
\usepackage{bm}

\begin{document}

\title{Vortex-induced negative magnetoresistance and peak effect in narrow superconducting films}

\author{D.Y. Vodolazov}
\email{vodolazov@ipmras.ru} \affiliation{Institute for Physics of
Microstructures, Russian Academy of Sciences, 603950, Nizhny Novgorod, GSP-105, Russia}

\date{\today}


\begin{abstract}

In framework of Ginzburg-Landau model it is shown that narrow superconducting film with width $w \simeq 3-8 \xi(T)$ ($\xi(T)$ is a temperature
dependent coherence length) exhibits unusual transport properties. In the absence of bulk pinning its critical current $I_c$ nonmonotonically
depends on perpendicular magnetic field H and has one minima (dip) and one maxima (peak) at some magnetic fields. At currents $I \ll I_c(H)$
the finite magnetoresistance $R(H)$ of such a samples due to thermo-activated vortex hopping via edge barriers also shows both local maxima(peak)
and minima(dip) nearly at the same magnetic fields. In narrower films such an effect is absent due to absence of the vortices and in wider films
the effect is weaker due to increased vortex-vortex interaction. Finite length of the film produces additional periodic variation in both $I_c(H)$
and $R(H)$ because of discrete change in the number of the vortices, which is superimposed on the above mentioned nonmonotonic dependence.
The obtained results are directly related to many experiments on narrow superconducting films/bridges where such a nonmonotonic dependencies
$I_c(H)$ and $R(H)$ were observed.

\end{abstract}

\maketitle

\section{Introduction}

It is well known that mesoscopic superconductors (with lateral and transverse sizes comparable
with temperature dependent coherence length $\xi(T)$) have transport characteristics (critical current $I_c$ and resistance R) which nonmonotonically depends on applied magnetic field H.
Probably the most familiar example is a Little-Parks effect when the resistance of the hollow superconducting cylinder (so called double connected system) varies periodically with H because of the change in vorticity \cite{Little}. Variations of the critical current  in double connected system (superconducting ring) are found in Ref. \cite{Gurtovoi,Michotte} while the same effect in single connected superconducting samples (squares, triangles, etc.) is experimentally observed in Refs. \cite{Vodolazov5,Schildermans,Falk,Aladyshkin}. Physically, both in single connected and double connected geometries effect is connected with adding to the screening current $j_{scr}$, induced by the external magnetic field, the current which flow around the vortex (or current created by the fluxoid in the ring and hollow cylinder). These currents cancel each other (fully or partially) and it is reflected in periodic variation of $I_c$ and R with H. In large scale system (with sizes $\gg \xi$) the effect practically disappear (amplitude of variation of $I_c$ and $R \to 0$) because current induced by the single vortex decays fast far from the vortex core.

In this paper we show that thin narrow superconducting film with width $3\xi \lesssim w \lesssim 8\xi \ll \Lambda$ ($\Lambda=2\lambda^2/d$, where $\lambda$ is the the London penetration depth and $d<\lambda$ is a thickness of the film) placed in perpendicular magnetic field has
nonmonotonic $I_c(H)$ and $R(H)$ although its length may go to infinity. As compared with the mesoscopic samples the effect is mainly connected not with a change in the number of the vortices but with the appearance of the vortex row in the film when the magnetic field increases.

Let us first discuss what is the minimal width of the film in which this effect may exist. In Ref. \cite{Saint-James} it is shown (in framework of Ginzburg-Landau model) that vortices may appear only in the films with width $w^* \gtrsim 1.8 \xi(T)$  at finite magnetic field and they do not penetrate to the narrower film at any magnetic field smaller than $H_c$ ($H_c$ is a critical field at which the superconductivity vanishes and $H_c$ is equal to third critical field $H_{c3}$ when $w\gg \xi$ and $H_c \sim 1/w$ for the films with $w \ll \xi$). Sometimes in the literature it is used the different critical width $\widetilde{w}^*\simeq 4.4 \xi$ which follows from numerical calculations for the superconducting bridge attached to a bulk electrodes \cite{Likharev1}. Note that last result is found in case when $H=0$ and state with a vortex sitting in the center of the film (when $I\to 0$) is a {\it saddle point} state. In Refs. \cite{Qiu,Vodolazov3} it is argued that energy of such a vortex state $U_{vortex}$ practically coincides with the energy of the Langer-Ambegaokar $U_{LA}$ saddle point state \cite{Langer} (in this state the order parameter vanishes along
the line connecting opposite edges of the film - see inset in Fig. 1 of Ref. \cite{Vodolazov3}) when $w \lesssim 4.4 \xi$. In wider films $U_{vortex}<U_{LA}$ and it takes less energy to create a vortex than LA state at $I\ll I_{dep}$ (but still there is a range of the currents very close to depairing current $I_{dep}$ where $U_{vortex}>U_{LA}$ even for films with $w \gg 4.4 \xi$ \cite{Vodolazov3}).

When $w>w^*$ it is energetically favorable to have vortices in the {\it ground} state of the film at magnetic fields $H_{c1}<H<H_c$ but as a {\it metastable} states the vortices also could exist at lower fields $H_0<H<H_{c1}$ due to finite energy barrier for vortex exit (it originates from trapping of the vortex by the screening current $j_{scr}(H)$). In the the London model magnetic fields $H_{c1}$ and $H_0$ were calculated in several works \cite{Abrikosov,Steijic,Clem} and for set of widths they were calculated numerically in the Ginzburg-Landau (GL) approach in Ref.  \cite{Sanchez}.

The increase of $I_c$ with appearance in the narrow film one disperse vortex row (with intervortex distance $a \gg w$) was theoretically predicted by Shmidt \cite{Shmidt} more than 40 years ago in framework of London model. Physically this effect was explained by increased trapping of the vortices by $j_{scr}(H)$ when magnetic field increases and necessity to increase transport current to overcome this force. In Ref. \cite{Shmidt} it was argued that after reaching maximal value the critical current should decay at larger magnetic fields due to suppression of superconductivity by magnetic field and hence it should be peak in dependence $I_c(H)$ (see curve $I_c^a$ in Fig. 1).

Much later in Refs. \cite{Mawatari,Carneiro} it was found that entrance of second, third and subsequent vortex rows to the film leads to additional dips (and peaks) in dependence $I_c(H)$. In case of relatively wide film (in the sense that $\xi \ll w < \Lambda$) one may use continuous approach with coordinate dependent vortex density when the number of vortex rows is large. This approach was utilized by Maksimova \cite{Maksimova} and it was predicted monotonic decay of critical current in increasing magnetic fields.

Experimentally dependence $I_c(H)$ was studied in various narrow superconducting films. One dip/peak in $I_c(H)$ was found for Nb film with $w\sim 4-5 \xi$ \cite{Ichkitidze}, several dips/peaks were present for the film with $w\sim 7-10 \xi$ \cite{Yamashita} and no dips and a monotonic $I_c(H)$ was observed for Nb and NbN films with $w \gg \xi$ in Refs. \cite{Steijic,Gershenzon,Henrich}. It is important that in these experiments the effect of bulk pinning was negligible at low magnetic fields and dependence $I_c(H)$
was governed only by edge/surface barrier effect (impact of bulk pinning in the film with
edge barrier for vortex entry/exit was discussed in Refs. \cite{Plourde,Vodolazov1} and analytically it was studied in Refs. \cite{Maksimova2,Elistratov}).
\begin{figure}[hbtp]
\includegraphics[width=0.53\textwidth]{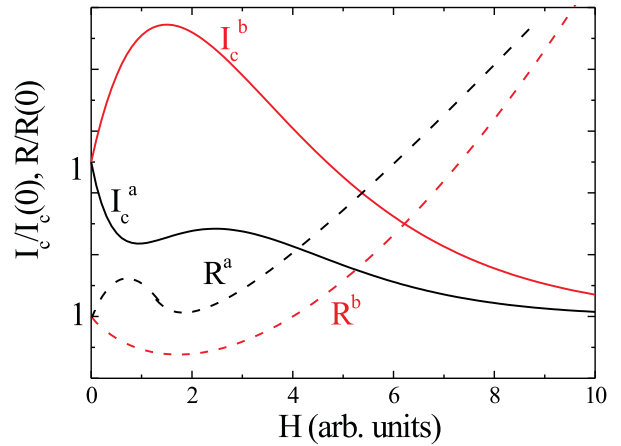}
\caption{Sketch of two types nonmonotonic dependence $I_c(H)$ (solid curves) and related with
them $R(H)$(dashed curves) observed in many experiments on narrow superconducting films.}
\end{figure}

Note, that enhancement of $I_c$ with increase of H was also observed in narrow superconducting wires with width $w \lesssim \xi$ \cite{Rogachev,Tian} where one cannot expect effect of vortices. Qualitatively the dependence $I_c(H)$ had a form which is different from above mentioned dependence - see curve $I_c^b$ in Fig. 1. We believe that enhancement of $I_c$  found in Refs. \cite{Rogachev,Tian} has different origin (for discussion of this behavior see Refs. \cite{Wei,Kharitonov1,Vodolazov2}) and it is not connected with an appearance of the vortices in the film/wire.

When $I_c$ is a nonmonotonic function of $H$ one may expect that resistance also changes nonmonotonically with $H$. Indeed, at current larger than $I_c(H)$ finite resistance appears due to vortex motion and just above $I_c(H)$ one may write $R \sim (I-I_c(H))$ (assuming that there is not voltage jump at $I=I_c$). Therefore if one fixes current $I$ and changes $I_c$ by applying magnetic field then variations in $I_c(H)$ will be directly reflected in variations of $R(H)$. This problem for narrow film was numerically studied in recent work \cite{Berdiyorov} (using time-dependent Ginzburg-Landau equation) and authors found nonmonotonic $R(H)$.

More complicated question is the finite resistance below $I_c$. If one uses concept of the energy barrier $U$ for the vortex entry/exit to/from the superconductor then, by definition, these barriers vanish at $I=I_c(H)$. Due to thermoactivation the vortex have finite probability $P$ to enter/exit the superconductor even when the height of the barriers is finite (at $I<I_c(H)$) and passage of the vortex through the superconductor leads to the voltage pulse and finite resistance. In the simplest model one may write that near the critical current $U\sim U_0(1-I/I_c)^m$ (for example $m\simeq 1$ for narrow film at zero magnetic field \cite{Vodolazov3}) and because $P\sim exp(-U/k_BT)$ the resistance is proportional to $\sim exp(-U_0(1-I/I_c)^m/k_BT)$. Therefore variations of $I_c$ in magnetic field should be reflected in variations of R even at $I < I_c$ (stress here that $I_c$ in above discussion is a theoretical critical current in the absence of fluctuations).

Experimentally nonmonotonic (negative) magnetoresistance in narrow superconducting films/wires was observed in many experiments \cite{Parks,Herzog,Johansson,Patel,Wang,Cordoba,Xiong,Rogachev,Tian1,Tian2,Chen1,Chen2,Zgirski,Gardner}.
As in the case with $I_c(H)$ one can distinguish two types of dependence $R(H)$. In one set of experiments \cite{Xiong,Rogachev,Tian1,Tian2,Chen1,Chen2,Zgirski,Gardner}
dependence R(H) had a dip at low magnetic fields and than resistance reached
normal state value at large H (see curve $R^b$ in Fig. 1). Such a behavior was mainly observed in
quasi-1D films/wires with width $w \lesssim \xi$ where no vortices can exist. At the moment there are several theories \cite{Wei,Pesin,Fu,Vodolazov2,Arutyunov,Kharitonov2}
which explain this effect by different mechanisms (for comparison of theoretical models
see Refs. \cite{Vodolazov2,Gardner}).

In another set of experiments \cite{Parks,Herzog,Johansson,Patel,Wang,Cordoba}
$R(H)$ had qualitatively different behavior. At weak magnetic fields first there was a peak in $R(H)$ which was followed by the dip
(see curve $R^a$ in Fig. 1). Besides, as in case of corresponding $I_c(H)$, there could be several dips/peaks in R(H) \cite{Parks,Johansson,Patel,Wang}.
Sometimes, the oscillations of R(H) with much smaller amplitude and shorter period could be superimposed on this nonmonotonic behavior
\cite{Johansson,Patel} and they were related to the change by one in the number of the vortices in the film \cite{Johansson}.

These experiments motivate us to calculate dependencies $I_c(H)$ and $R(H)$ for narrow films in wide range of widths. Contrary to previous theoretical
works on this subject we use Ginzburg-Landau approach because it takes into account suppression of the superconducting order parameter
by the screening/transport current (which is important when the critical current is close to depairing current or magnetic field is close to $H_c$)
and effect of the finite-size vortex core which are absent in the the London model and which are important from quantitative point of view.
Besides the GL model automatically correctly takes into account vortex-vortex interaction in the presence of edges (via boundary conditions
for superconducting order parameter) and resolves the question about stability of static vortex configurations in the film with transport current.
Previously $I_c(H)$ was already calculated in the GL model for narrow film (for restricted set of widths) and a dip/peak in dependence $I_c(H)$
was found in Refs. \cite{Vodolazov4,Gladilin} but its origin was not studied.

To calculate the magnetoresistance $R(H)$ we find the energy barriers for the vortex entry $U_{en}$ and vortex exit $U_{ex}$ both in the presence
and in the absence of the vortices in the film in the limit when $I\to 0$. For this purpose we find the solution of the Ginzburg-Landau equation
which corresponds to the saddle point state. The magnetoresistance is estimated by using the Arrhenius law $R(H)\sim exp(-U_{max}/k_BT)$, where
$U_{max}$ is a maximal energy barrier at given magnetic field $U_{max}(H)=max\{U_{en},U_{ex}\}$. Comparison of our results with existing experiments
showed good qualitative and sometimes quantitative agreement. The possible reasons for quantitative disagreement are discussed.

The paper is organized as follows. In section II we present the results for $I_c(H)$ and compare them with existing theories and experiments.
In section III we present results for $R(H)$ and compare them with the experiments. In section IV we conclude our main results.

\section{Dependence of the critical current on magnetic field}

\subsection{Model}

In numerical calculations we mainly use the length of the film $L=40 \xi$ and vary the width from $w=2\xi$ up to 20 $\xi$. At the ends of the film we apply normal metal-superconductor (NS) boundary conditions to inject the current to the superconductor. To avoid effect of these NS contacts on the vortex distribution and stability of the superconducting state we locally enhance critical temperature (on the distance $2.5 \xi$) near the ends. It leads to enhanced order parameter near the ends, which partially mimics the effect of bulk leads to which wire/film/bridge is usually attached in the experiment. We check that these places lose the superconducting properties at larger currents than the main part of the film.

The critical current is determined as a current at which vortex motion starts (without fluctuations) and voltage drop across the central part of the film becomes nonzero. Fluctuations (if they are strong enough) may provide switching of the superconductor to the resistive state at $I<I_c$ but because their probability is roughly proportional to $exp(-U_0(I_c(H)-I)^m)/k_BT$ (see discussion in Introduction) one may expect that $I_c(H)$ in the presence of fluctuations follows $I_c(H)$ in the absence of fluctuations (if $U_0/k_BT \gg 1$ and these fluctuations are relatively rear events).

In the model we assume that the London penetration depth $\lambda$ is much larger than the width of the film and hence one can neglect magnetic field which is induced by
the transport and screening currents. It considerably simplifies the calculations because we
have to solve only 2D Ginzburg-Landau equation for the superconducting order parameter $\Delta=|\Delta|exp(i \phi)$
\begin{eqnarray}
\frac{\pi\hbar}{8k_BT_c}\left( \frac{\partial}{\partial
t}+2ie\varphi \right)\Delta=
\\ \nonumber
\xi_{GL}^2\left(\nabla-i\frac{2eA}{\hbar c}\right)^2+\left(1-\frac{T}{T_c}-\frac{|\Delta|^2}{\Delta_{GL}^2}\right)\Delta.
\end{eqnarray}

In Eq. (1) $\xi_{GL}^2=\pi\hbar D/8k_BT_c$, $\Delta_{GL}^2=8\pi^2(k_BT_c)^2/7\zeta(3)$ and $D$ is a diffusion coefficient.
Vector potential ${\bf A}$ has only one component ${\bf A}=(0,Hx,0)$.
Equation for the electrostatic potential $\varphi$ follows from condition $div j=0$ and one obtains
\begin{equation}
\frac{\partial^2\varphi}{\partial x^2}+\frac{\partial^2\varphi}{\partial y^2}=\rho_n div j_s,
\end{equation}
where $\rho_n$ is a normal state resistivity and $j_s$ is a superconducting current density.

Eq. (1) is strictly valid only for gapless superconductors but we use it here not to study the dynamics of $\Delta$ but to find the current at which the stationary superconducting state (described by Eq. (1) with zero left hand side (LHS)) becomes unstable. Eq. (1) also provides the convenient way of finding stationary state (if it exists at given current and magnetic field) starting from initial condition with $|\Delta|(x,y)=\Delta_{GL}(1-T/T_c)^{1/2} $ and ending numerical calculations when LHS of Eq. (1) goes to zero.
\begin{figure}[hbtp]
\includegraphics[width=0.48\textwidth]{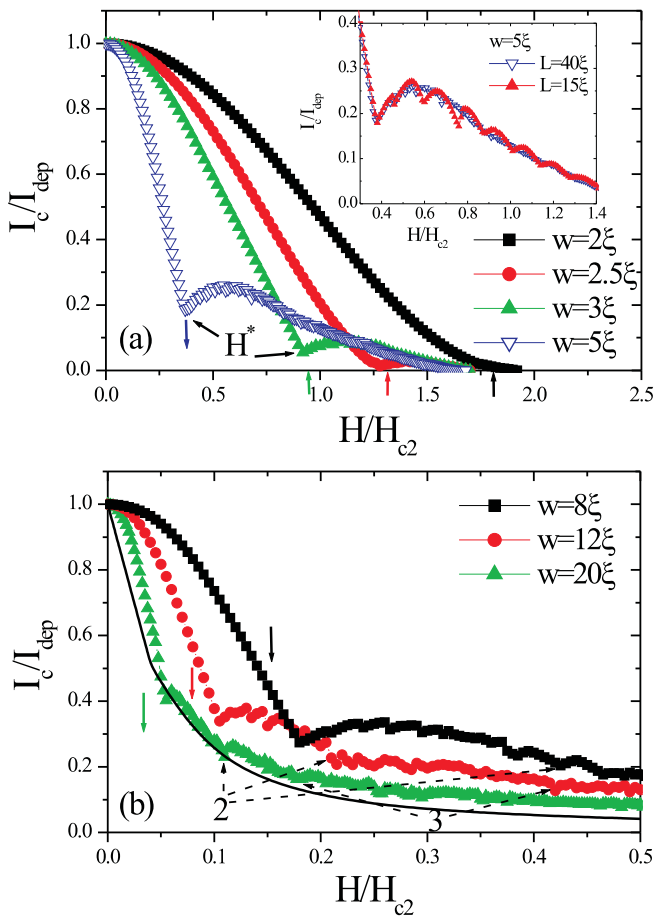}
\caption{Dependence $I_c(H)$ for films with different widths and length $L=40 \xi$. Numbers and dashed arrows in (b)
indicate the magnetic fields when second and third vortex rows appear in the film.
Color arrows indicate the positions of $H_{c1}$ for corresponding films. For relatively wide
films $w\gtrsim 3 \xi$ the critical current goes to zero at third critical magnetic field $H_{c3}\simeq 1.7 H_{c2}$ when surface superconductivity vanishes. Black solid curve in (b) corresponds to theoretical $I_c(H)$ which follows from the London model (see Eqs. (23,37) in Ref. \cite{Maksimova} or Eqs. (4,6) in \cite{Plourde}) for film with $w=20 \xi$. Inset in (a)
demonstrates the evolution of $I_c(H)$ when the length of the film decreases.}
\end{figure}

\subsection{Results}

In Fig. 2 we present calculated $I_c(H)$ for the films with different widths. When $w \lesssim 6 \xi$ in the film can exist only one vortex row at large H and there is one noticeable dip at $H=H^* \sim H_{c1}$ and one peak in dependence $I_c(H)$. For wider films more than one vortex row may appear in the film (see dashed arrows in Fig. 2(b)) and each time when new vortex row is nucleated in the film it leads to additional dips (and peaks) in dependence $I_c(H)$ but their amplitude is much smaller than for narrower films.

One can also notice short period oscillations of $I_c$ well visible for relatively wide films at low magnetic fields (when in the film exist only one vortex row - see Fig. 2(b)) and for shorter films (see inset in Fig. 2a). Their period $\Delta H$ depends on $w$ and
changes roughly from $\Delta H \simeq 1.3 \Phi_0/wL$ for film with $w=3 \xi$ up to $\Delta H \simeq 1.9 \Phi_0/wL$ for film with $w=20 \xi$ ($\Phi_0$ is a magnetic flux quantum). These oscillations are connected with change in the number of the vortices in the film by one and they are reminiscent of Fraunhofer-like oscillations of $I_c$ in wide Josephson junction when number of Josephson vortices changes by one. Similar oscillations were experimentally observed in mesoscopic single-connected superconductors \cite{Vodolazov5,Schildermans,Falk,Aladyshkin}.

It is interesting to note that qualitatively our results for evolution of $I_c(H)$ with increase of $w$ resemble $I_c(H)$ for long diffusive Josephson junction of finite width (compare Fig. 2(a,b) with Fig. 3 in Ref. \cite{Cuevas}). The difference is that in Josephson junction the number of the {\it vortices} changes by one (and it leads to appearance of new dips/peaks) while in case of film the number of the {\it vortex rows} changes by one. The reason for such a difference is clear from Fig. 3. In the Josephson junction there is only one vortex row which is {\it perpendicular} to the current, while in the film it could be several rows which are {\it parallel} to direction of current flow. Besides, in the superconducitng film the number of the vortices in the row may vary with H, which provides additional source of oscillations of $I_c(H)$ (mentioned in previous paragraph). And last quantitative difference
is that for Josephson junction in the dips $I_c$ goes to zero (see for example Ref. \cite{Cuevas}), while for narrow films it is finite there.
\begin{figure}[hbtp]
\includegraphics[width=0.35\textwidth]{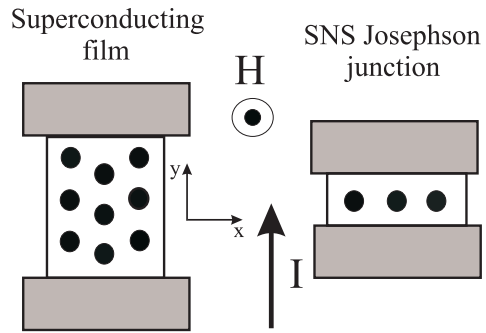}
\caption{Sketch of vortex distribution in superconducting film and Josephson junction with
finite length and width placed in perpendicular magnetic field.}
\end{figure}

Let us now discuss why nonmonotonic behavior of $I_c(H)$ is the strongest one for relatively narrow films with $w\simeq 3-8 \xi$ (we consider now case of long films when short period oscillations have very small amplitude and may be discarded). In Fig. 4 we plot current density distribution across the film at nonzero transport current and various H which follows from the London model
in the vortex free state. This current density is a sum of the transport current density $j_{tr}=I/wd$ and screening current density $j_{scr}(x)=-cHx/4\pi\lambda^2$. When sum $j_{tr}+j_{scr}$ on the left edge equals to depairing current density $j_{dep}$ the superconducting state becomes unstable and vortices enter the film (in terms of energy barriers the barrier for vortex entry goes to zero on the left edge at this condition). These vortices may freely pass the film when $j_{tr}+j_{scr}>0$ everywhere in the film (lines 1,2 in Fig. 4) because the Lorentz force $F_L=[j,\Phi_0]/c$ acting on the vortex directs to the right edge and force from the
vortex images $|F_{image}|$ is smaller than $|F_L|$ in the left half of the film. Keeping full current density at the left edge equal to $j_{dep}$ and varying H one can easily find linear decay $I_c(H)=I_c(0)(1-H/H_s)$ \cite{Maksimova} at fields $0<H<H_s/2$, where $I_c(0)=j_{dep}wd$ and $H_s$ is a superheating magnetic field (at this field the surface barrier for vortex entry goes to zero at I=0).
\begin{figure}[hbtp]
\includegraphics[width=0.45\textwidth]{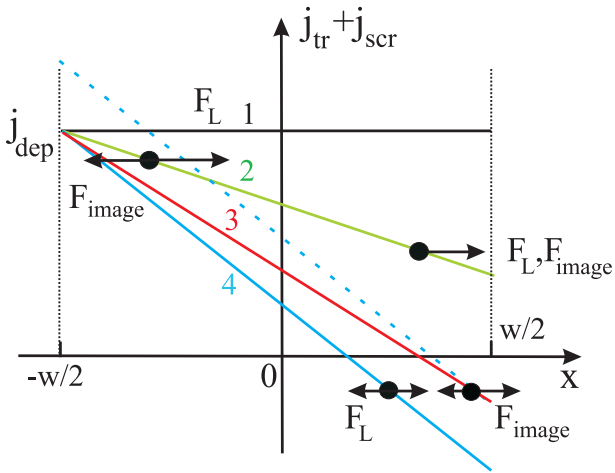}
\caption{Sketch of current density distribution in narrow superconducting film. Black spot
indicates the position of vortex and forces which act on it. Line 1 corresponds to field $H^1=0$ and line 2 to $H^2<H_s/2$. Lines 3 and 4 correspond to $H^4>H^3>H_s/2$. When $j_{tr}+j_{scr}$ changes the sign near the right edge of the film the vortex stops in the point where $|F_L|=|F_{image}|$ (line 4). One needs to increase the current (dashed line) to move the vortex in the point where $|F_{image}|>|F_L|$ and vortex can exit the film. Area under the lines determines the transport current in the film.}
\end{figure}

At field $H>H_s/2$ the sum $j_{tr}+j_{scr}$ changes the sign close to the right edge and vortex would stop near this point because Lorentz force changes the sign there. But if this point is not far from the right edge the force from vortex images is larger than the Lorentz force and
vortex is able to exit the film (line 3 in Fig. 4). At larger magnetic field the vortex
already cannot leave the film (line 4 in Fig. 4) and one has to increase the current in the system to shift the vortex closer to the right edge in the point where $|F_{image}|>|F_L|$ (dashed line in Fig. 4). From Fig. 4 one can see that the area under dashed line is larger than under line 3 and therefore the critical current is also larger. This consideration gives the physical background for increase of $I_c$ at fields $H \gtrsim H_s/2$ (this explanation of peak effect is alternative to one present in Ref. \cite{Shmidt}).

In above simple picture we use single vortex approach (as in Ref. \cite{Shmidt}). But when the barrier for vortex entry is suppressed one will have not a single vortex but a vortex row with a period $a$ which depends on applied magnetic field (see Figs. 5,6) (in case $I=0$ dependence $a(H)$ for film with $w=5 \xi$ is calculated in Ref. \cite{Sanchez}).
\begin{figure}[hbtp]
\includegraphics[width=0.42\textwidth]{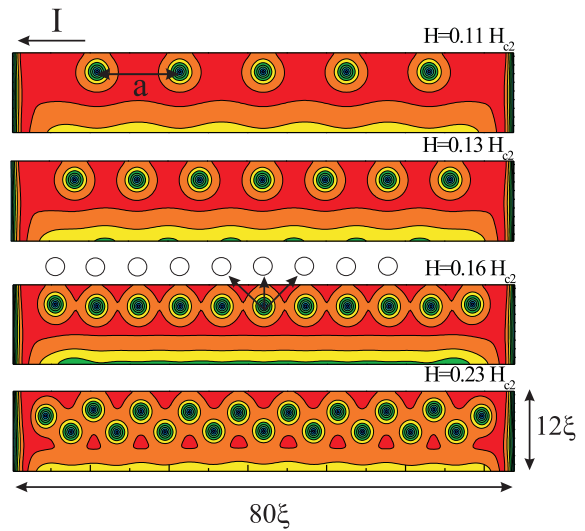}
\caption{Distribution of $|\Delta|$ in the film with $w=12 \xi$ and $L=80 \xi$ at different magnetic fields and currents just below $I_c(H)$.
One can see that already at $H\gtrsim H^*=0.105 H_{c2}$
intervortex distance is smaller than width of the film. The empty circles qualitatively demonstrate the position of the nearest vortex images
and increased attraction to the edge due to images of adjacent vortices.}
\end{figure}
\begin{figure}[hbtp]
\includegraphics[width=0.42\textwidth]{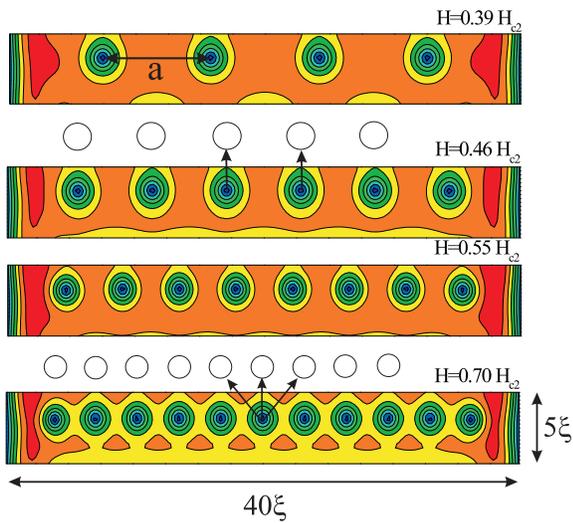}
\caption{Distribution of $|\Delta|$ in the film with $w=5 \xi$ and $L=40 \xi$ at different magnetic fields and currents just below $I_c(H)$.
The empty circles qualitatively demonstrate the position of the nearest vortex
images. At relatively low $H$ (when $a>w$) main attraction to the edge comes from own image of the vortex, while at large H (when $a<w$)
in addition there is noticeable attraction from images of adjacent vortices placed on distance less than w.}
\end{figure}

When the vortex row enters the film it decreases the current density on the edge where it enters because the current which flows around the vortices $j_{vort}$ has a different sign with current $j_{scr}+j_{tr}$. The vortex row stops at distance $r\sim w $ from left edge (left edge in Fig. 4 corresponds to bottom edge in Figs. 5,6) and hence reduction of $j$ at that edge will be weaker for wider films $w \gg \xi$ in comparison with relatively narrow film where $w \sim \xi$ (see Fig. 6).
Therefore it is possible to have a situation when with increase of the transport current the new vortices enter the film (when $j_{scr}+j_{tr}+j_{vort} \geq j_{dep}$) before already existing vortex row exit the film and it launches the continuous vortex motion and resistive state.
Our numerical calculations show that for relatively wide films the resistive state starts according to this scenario (at least when in the film exist one vortex row). Then it becomes clear that the wider the film the less one should increase $I$ to create new vortex row in the film and
it explains weak peak effect in films with $w \gg \xi$. In some respect the situation is similar to mesoscopic samples with size (several $\xi \times$ several $\xi$) where oscillations of $I_c$ are connected with compensation of the $j_{scr}+j_{tr}$ by $j_{vort}$ and amplitude of oscillations
of $I_c(H)$ becomes small with increase the length and width of the superconductor.

In films with $w \lesssim 10 \xi$ our numerical calculations show that resistive state at $H>H^*$ starts from exit of the vortices, subsequent entry of new vortices and so on. We believe that in such a films $I_c$ grows up to the magnetic field at which the intervortex distance becomes about the width of the film. At larger fields when $a<w$ each vortex in the row stronger interacts
with adjacent vortices and their images (at $a>w$ the vortex-vortex interaction decays exponentially with distance between vortices \cite{Shmidt,Steijic}) which clearly enhances the attraction of the vortices to the nearest edge (see Fig. 6). The same effect exists in relatively wide films where even small increase of $H$ above $H^*$ leads to $a<w$ (see Fig. 5). In both cases the enhanced trapping of vortex due to $j_{scr}(H)$ is compensated by increased interaction with the edge of the film due to smaller $a$ and when $a(H)<w$ critical current decreases with increase of H. At large magnetic fields $H\simeq H_c$ additional decay of $I_c$ comes also from suppression of $|\Delta|$.

We check that the peak effect is robust with respect of presence of the localized edge defects and suppression of the superconductivity along the edges. Last effect is discussed in Ref. \cite{Il'in} to explain the lowering $T_c$ of the narrow films with decreasing their width. To model the localized edge defects we locally suppresses $T_c$ in the semicircle with radius $\xi$ placed at each edge (see inset in Fig. 7) while suppression of superconductivity along whole edge of the film we model by reduction of $T_c$ on distance of $\xi/4$ near the edges (see inset in Fig. 7 and inset in Fig. 15). In both cases this procedure leads to local suppression of $|\Delta|$ in the defect region and far from it (due to proximity effect - see inset in Fig. 7).

In Fig. 7 we present $I_c(H)$ for both types of defects (lower value of $\Delta_{edge}$
corresponds to lower value of local $T_c$). Suppression of $T_s$ along edges shifts the position of the dip to larger fields which is explained by reduction of the effective 'superconducting' width  of the film (region which possesses the superconducting properties - see
inset in Fig. 15). The localized edge defects lead to appearance of two peaks (see curve with empty squares in Fig. 7). First peak is connected with single vortex localized near one of the edge defects (in the region between the defect and NS boundary) and second with a vortex row which appears at larger H. If one decreases $T_c$ in the localized defect (which leads to locally smaller value of $|\Delta|$) dependence $I_c(H)$ becomes irregular (not shown here) and it is hard to notice one pronounced peak. This could be explained by presence of various widths (near localized defect the 'superconducting' width of the film is effectively smaller) and peaks, which appear at different magnetic fields, interfere each other. It shows that the peak effect, in some respect, is collective effect and to be observable one has to have relatively small variations
of physical properties along the film.
\begin{figure}[hbtp]
\includegraphics[width=0.52\textwidth]{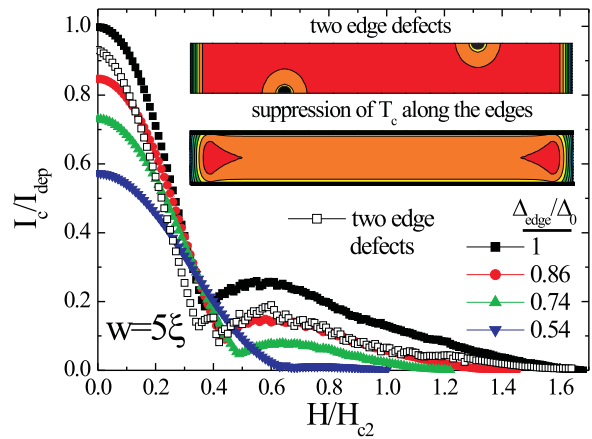}
\caption{Dependence $I_c(H)$ for narrow film ($w=5\xi$) with suppression of $T_c$ along the edge leading to smaller value of the order parameter at the edge $\Delta_{edge}$ (it is normalized by $\Delta_0=\Delta_{GL}(1-T/T_c)^{1/2}$). Curve with empty squares corresponds to the film with two localized edge defects (with chosen $T_c^{loc}$ the minimum value of $|\Delta|$ in localized
defect equals to 0.76 $\Delta_0$). In the inset we present distribution of $|\Delta|$ calculated at $H=0$ and $I=0$ for two types of suppression of $T_c$ (regions with locally suppressed
$T_c$ are marked in black color).}
\end{figure}

\subsection{Comparison with the London model}

In this subsection we compare our numerical results with ones found in the London model \cite{Shmidt,Mawatari,Carneiro,Maksimova}.

From Eqs. (18,20) of Ref. \cite{Shmidt} one may find position of the dip $H^*\simeq (H_s+H_0)/2$. It gives $H^*\simeq H_s/2$ for wide films $w\gg \xi$ ($H_0\ll H_s$) and $H^*\simeq H_0\sim H_{c1}\sim H_s$ when one approaches critical width $w^*$ (where $H_0\sim H_{c1}\sim H_s$)
which qualitatively coincides with our numerical results. The amplitude of peak and its position could not be find from single vortex approach used in Ref. \cite{Shmidt}.

In Refs. \cite{Mawatari,Carneiro} the vortex-vortex interaction is taken into account in the framework of the the London model and calculated $I_c(H)$ demonstrates much higher peaks than our $I_c(H)$ for comparable width of the film ($w=25 \xi$) - see Fig. 2
in Ref. \cite{Mawatari} and Fig. 9 in Ref. \cite{Carneiro}. Another quantitative difference is in the ratio of critical currents at $H=0$ and at $H=H^*$. For moderately wide films $w \gtrsim 10 \xi$ this ratio is about 2 in the GL model (see Fig. 2(b)) and in single
vortex approach (it follows from Eqs. (18,20) in Ref. \cite{Shmidt}) while in Refs. \cite{Mawatari,Carneiro} it was found $I_c(0)/I_c^{dip}\sim 10-20$.

Our $I_c(H)$ for the widest film ($w=20 \xi$) agrees semi-quantitatively with the result of Maksimova \cite{Maksimova} (see black curve in Fig. 2(b)) for which we use $H_s=0.083H_{c2}$ found from the GL model. The quantitative differences (nonlinear versus linear $I_c(H)$ at low H and larger values of $I_c(H)$ in the GL model at high H) are well explained by limitations of the London model.
The nonlinear drop of $I_c$ at low magnetic fields comes from suppression of $|\Delta|$ by the transport current in the GL model which is most noticeable when $I_c\sim I_{dep}$. This effect was previously discussed \cite{Maksimova2,Andrackii} and experimentally confirmed in Ref. \cite{Andrackii} for narrow Sn bridges. If due to some reason (for example presence of localized defects) $I_c$ at zero magnetic field drops well below $I_{dep}$ one may recover
linear decay of $I_c$ at low magnetic fields even in the GL model. The larger values of $I_c$ in the GL model at high H comes from the edge vortex free layers with width about $\xi$ (they provide finite $I_c$ up to $H=H_{c3}$) and this effect cannot be catched by the London model where formally $\xi \to 0$.

\subsection{Comparison with the experiments}

Experiments on narrow films with $\xi \ll w <\Lambda$ did not reveal presence of dips/peaks in $I_c(H)$ \cite{Gershenzon,Steijic,Henrich} (the narrowest studied film had a width $w\simeq 14 \xi$ \cite{Gershenzon}). Direct comparison with an analytical dependence following from the London model \cite{Maksimova} showed good quantitative agreement between theory and experiment \cite{Henrich} at low magnetic fields where $I_c$ linearly drops with $H$.

Narrow films with width $w \simeq 3-7 \xi$ were experimentally studied in Refs. \cite{Ichkitidze,Yamashita}. In both experiments $I_c(H)$ showed pronounced dips/peaks. Position of the first (single in Ref. \cite{Ichkitidze}) dip roughly follows $H=H^*\sim H_{c1}$ where dependence $H_{c1}(w)$ found in the GL model is shown in Fig. 8 ($H_{c1}$ is found from the condition that at $H=H_{c1}$ the energies of the film with one vortex and vortex free state are equal). Besides the ratio $I_c(0)/I_c^{dip}$ extracted from Fig. 1 of Ref. \cite{Ichkitidze} for the film with $w\simeq 4\xi$ (according to the table present in Ref. \cite{Ichkitidze}) is close to our value calculated for the film with $w=4-5 \xi$. It is more difficult
to make the quantitative comparison with the results of Ref. \cite{Yamashita} because of logarithmic scale of shown dependence $I_c(H)$ and no information for actual width of the film in units of $\xi(T)$ at given temperature but it is close to our results for the film with $w \sim 4-5 \xi(T)$.

\begin{figure}[hbtp]
\includegraphics[width=0.53\textwidth]{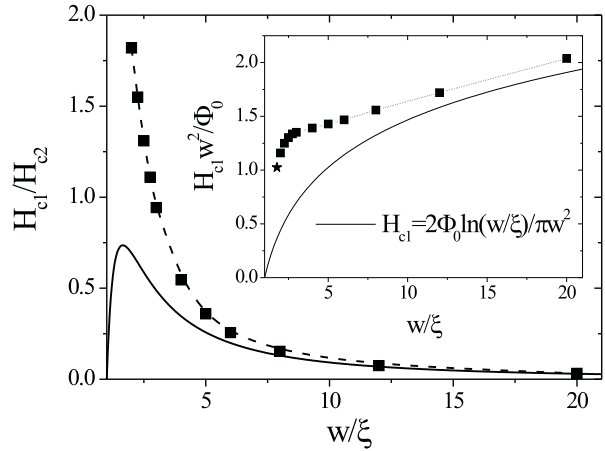}
\caption{Dependence of the first critical field $H_{c1}$ (squares) on the width of the film found from the numerical calculations in the GL model. Black solid line follows from the London model \cite{Shmidt,Steijic,Mawatari}. At $w = 2 \xi$ field $H_{c1}$ is
very close to the critical field $H_c$ and we did not study more narrow films. Star in the inset corresponds to $H_{c1}=H_c$ for the film with $w=w^*\simeq 1.8 \xi$ present in Ref. \cite{Saint-James}.}
\end{figure}

\section{Field dependent energy barrier and magnetoresistance}

In this section we calculate the field dependent energy barriers for vortex entry/exit to/from the narrow film when $I\to 0$. These results than are used to find the magnetoresistance of narrow films due to thermoactivated vortex hopping via these barriers.

\subsection{Model}

In Fig. 9 we illustrate hopping of the single vortex via energy barriers at different magnetic fields. At $H<H_0$ one needs to supply energy $U_{en}(H)$ to have a passage of the vortex across the film. Therefore at fields less than $H_0$ finite resistance is proportional to $exp(-U_{en}(H)/k_BT)$.

At fields larger than $H_0$ there is a local minimum in the dependence $U(x)$. Taking into account that vortex motion in the superconductors is strongly damped one may conclude that after overcoming entry energy barrier the energy of vortex will nearly follow the profile $U(x)$ and hence vortex stops in the local minimum of $U(x)$. To exit the film it should overcome the barrier $U_{ex}$ and
at the first sight the resistance at $H>H_0$ should be proportional to $exp(-(U_{en}(H)+U_{ex}(H))/k_BT)$ (which comes from product of probabilities to enter and to exit the film). But usage of the Arrhenius expression for estimation of vortex passage across the film implies that one should take into account not the sum of the barriers but the maximal barrier. Indeed, let us suppose for definiteness that $U_{en}>U_{ex}$ (as in Fig. 9). Due to finite temperature there is finite probability $P\sim exp(-\Delta U/k_BT) $ to deliver energy $\Delta U$ in every part of the superconductor in each moment of time. When a fluctuation with the energy $\Delta U=U_{en}$ occurs near the edge the vortex enters the film and than stops in the center. To exit the film it needs smaller energy $\Delta U=U_{ex}<U_{en}$ and probability of such an event is much larger $\sim exp(-U_{ex}/k_BT)\gg \exp(-U_{en}/k_BT)$ than for the vortex entry. Therefore the largest barrier creates some kind of bottleneck and it determines the rate of vortex hopping across the entire film.

In above consideration one implicitly assumes that the pre-exponential factor gives the small contribution to the probability for vortex entry/exit. It is the case when $\Delta U/k_BT \gg 1$. In opposite case this simple approach becomes invalid and one has to calculate the pre-exponential factor and its dependence on $\Delta U$.

\begin{figure}[hbtp]
\includegraphics[width=0.45\textwidth]{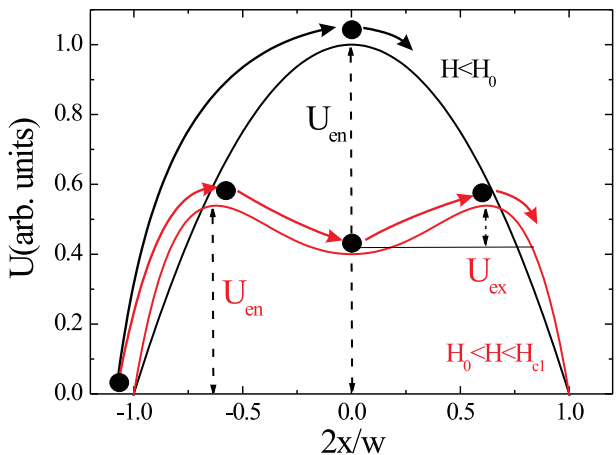}
\caption{Sketch of the energy profile of the probe vortex placed in point x in the narrow film at magnetic fields $H<H_{c1}$.}
\end{figure}

In the vortex free (Meissner) state $U_{en}(H)>U_{ex}(H)$ up to the field $H_{c1}$ and probability
for vortex entry $P_{en}\sim exp(-U_{en}(H)/k_BT)$ is smaller than for vortex exit
$P_{ex}\sim exp(-U_{ex}(H)/k_BT)$. Therefore on average in time there is no vortices
in the film at $H<H_{c1}$ (here we assume that above probabilities are not extremely low and vortices cannot be frozen in the film for very long times) and one may use single vortex approach for calculation of the energy barriers.

At fields larger than $H_{c1}$ there are vortices in the film (because $P_{en}>P_{ex}$) and one has to take them into account. Again, we assume that to observe finite resistance in the experiment the probabilities $P_{en}$ and $P_{ex}$ should not be extremely low. Consequently for any magnetic field the number of the vortices in the film is defined from the balance $P_{en} \sim P_{ex}$ which coincides with a condition that the film is being in the ground state. This assumption considerably simplifies calculation of $U_{en}$ and $U_{ex}$ because one may consider only transitions from ground state to the nearest metastable state (where number of vortices is larger/smaller by one). But even in this case there are two possibilities for thermoactivated vortex travelling across the film - see Fig. 10. Numerical calculations of the energy barriers show that in case 1 the maximal energy barrier corresponds to the barrier for exit (at $H>H_{c1}$) while in case 2 the maximal energy barrier corresponds to the barrier for entry (at $H>H_{c1}$) and it is larger than $U_{ex}$ except near the magnetic fields at which the number of the vortices in the film changes by one in the ground state (see Fig. 11). Because of similar dependencies of $U_{max}(H)$ in both cases and mainly smaller value of $U_{max}$ in case 1 than in case 2 we calculate the energy barriers for vortex hopping marked as case 1 in Fig. 10.
\begin{figure}[hbtp]
\includegraphics[width=0.45\textwidth]{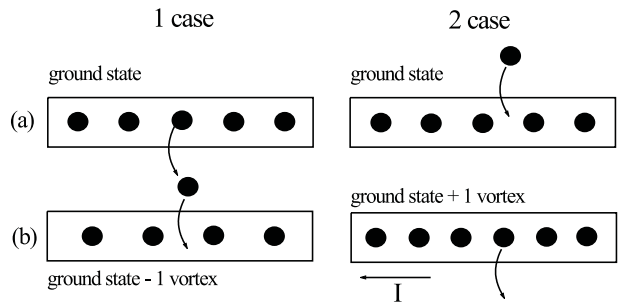}
\caption{Two scenarios of vortex passage through the film being in the ground state at $H>H_{c1}$. Case 1 has lower maximal energy barrier except near the magnetic fields where number of the vortices in the film changes by one in the ground state.}
\end{figure}

\begin{figure}[hbtp]
\includegraphics[width=0.45\textwidth]{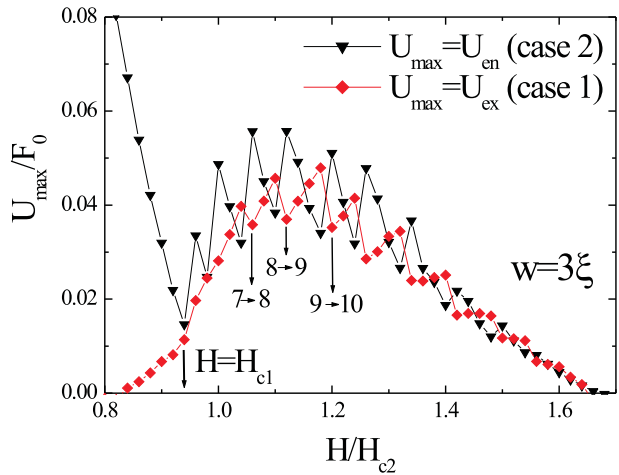}
\caption{Dependence of the maximal energy barrier (at $H>H_{c1}$) for vortex entry/exit on magnetic field for two scenarios of vortex passage through
the film (see Fig. 10). Numerics indicate the change in the number of the vorticies in the film at corresponding magnetic fields.
The width of the film $w=3\xi$ and the length is 40 $\xi$. The energy is normalized in units of $F_0=\Phi_0^2/8\pi^2\Lambda$.}
\end{figure}

To calculate the energy barriers for vortex entry/exit we numerically find solution of Ginzburg-Landau equation corresponding to the saddle point (SP) state \cite{Langer} at given value of magnetic field and number of vortices using the method of Ref. \cite{Vodolazov3}. For vortex free (Meissner) state we place the probe vortex along the central line of
the film (see Fig. 12(a,b)) and find profile $U(x)$ from which one can easily extract $U_{en}$ and $U_{ex}$. When the maximum of dependence $U(x)$ approaches the edge then instead of the vortex saddle point state contains the vortex nucleus (finite size region with partially suppressed $|\Delta|$ \cite{Vodolazov3,Schweigert}) sitting at the edge of the film. To find such a state
we fix magnitude of the order parameter in one point of the numerical grid near the edge and vary $|\Delta|$ in this point (keeping H constant) until such a state becomes nonstationary and the vortex enters the film \cite{Vodolazov3} (the same procedure is used for finding $U_{ex}$ when we displace the vortex to the edge). The example of such a state is shown in Fig. 12(c).
\begin{figure}[hbtp]
\includegraphics[width=0.48\textwidth]{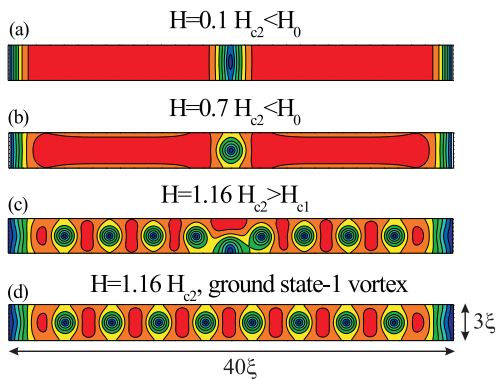}
\caption{(a-c) Distribution of $|\Delta|$ in saddle point state at different magnetic fields. In (d) we plot $|\Delta|$ in the metastable state which corresponds to state (b) of case 1 in Fig. 10.}
\end{figure}

By knowing both barriers one can calculate the magnetoresistance using expression
\begin{equation} 
R(H)=\nu exp(-U_{max}/k_BT).
\end{equation}
where $U_{max}=max\{U_{en},U_{ex}\}$. Eq. (3) contains prefactor $\nu$ which can be found only from solution of time-dependent problem
\cite{McCumber,Golubev}. Its calculation for quasi-1D superconducting wire in the limit when fluctuations are rear events ($U/k_BT \gg 1$)
showed that $\nu \sim (U/k_BT)^{1/2}$ \cite{McCumber,Golubev}. But when $U/k_BT \gg 1$ it is clear that the main dependence $R$ on $H$ comes
from the exponent. Besides when $H\to H_c$ and $U\to 0$ one should have normal state resistance $R_n$. Therefore for calculation (estimation)
of magnetoresistance we use the following semi-phenomenological expression
\begin{equation}
R(H)=R_nexp(-U_{max}/k_BT)
\end{equation}

\subsection{Results}

In Fig. 13 we present dependence of maximal energy barrier $U_{max}=max\{U_{en},U_{ex}\}$ on the applied magnetic field for the films with widths $w=2-5 \xi$ (the energy is normalized in units of $F_0=\Phi_0^2/8\pi^2\Lambda$). The maximal barrier corresponds to $U_{en}$ at $H<H_{c1}$ and to $U_{ex}$ at larger fields for vortex hopping marked as case 1 in Fig. 10. Similar to $I_c(H)$ there is a range of magnetic fields where $U_{max}$ increases with increase of H (and hence R decreases according to Eq. (4)). In the inset to Fig. 13 we show high field region for the film with $w=3 \xi$ where one can see that both barriers $U_{en}$ and $U_{ex}$ increase at $H>H_{c1}$. One can also notice short period oscillations of $U(H)$ (with practically the same period as for $I_c(H)$) which are connected with change in the number of the vortices.
Due to finite length of the film intervortex distance in the row changes discontinuously and it results in jumps of both $U_{en}$ and $U_{ex}$. When number of vortices is constant both barriers varies continuously: $U_{en}$ goes down and $U_{ex}$ goes up because $j_{scr}(H)$ gradually increases when magnetic field grows.
\begin{figure}[hbtp]
\includegraphics[width=0.45\textwidth]{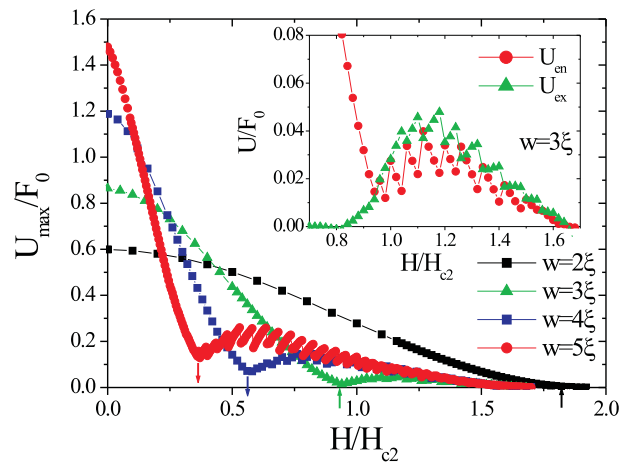}
\caption{Dependence of the maximal energy barrier for vortex entry/exit on magnetic field for films with different width. In the inset we plot both $U_{en}$ and $U_{ex}$. Color arrows indicate field $H_{c1}$ for given film. The length of the films is 40 $\xi$.}
\end{figure}

For films with $w \gtrsim 8 \xi$ relative increase in $U_{max}(H)$ is much smaller when for narrower films (compare Figs. 13 and 14). In such a films the intervortex distance becomes comparable with $w$ already at fields close to $H_{c1}$ (see Fig. 15) and enhanced trapping of the vortices by $j_{scr}$ is compensated by vortex-vortex repulsion and attraction by edges. In narrower films, $a(H)>w$ in relatively wide range of magnetic fields (see Fig. 6, where number of the vortices at $I\simeq I_c(H)$ is the same as at $I=0$, contrary to film with $w=12 \xi$) and $U_{ex}$ grows up to the field where $a(H) \lesssim w$ and at larger fields $U_{ex}$ decreases. In addition, in relatively narrow films in which $H_{c1}\sim H_{c2}$ the order parameter is strongly suppressed in the film at $H \lesssim H_{c1}$ and entrance of the vortex row increases $|\Delta|$ at the edge (due to compensation of $j_{scr}$ by $j_{vort}$). It provides increase of $U_{en}(H)$ in the film with $w=3\xi$ (see inset in Fig. 12) but for wider films (where $H_{c1}\ll H_{c2}$) this effect is weaker.
\begin{figure}[hbtp]
\includegraphics[width=0.52\textwidth]{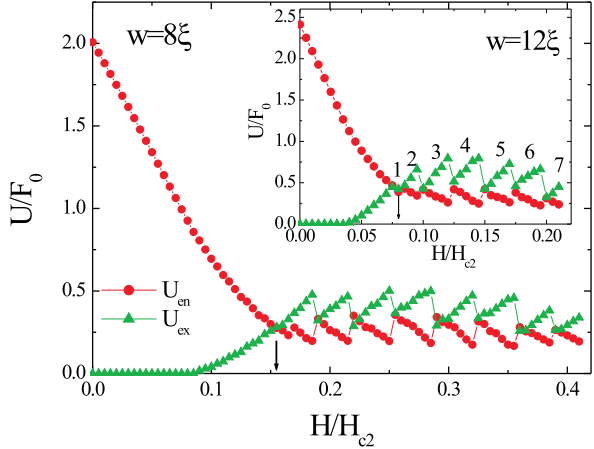}
\caption{Dependence of $U_{en}$ and $U_{ex}$ on magnetic field for the film with $w=8\xi$. In the inset the results for the film with $w=12 \xi$ are present (numerics show the number
of vortices at corresponding magnetic fields). The barriers are calculated up to the
field when the second vortex row appears in the film. Arrow indicate field $H_{c1}$. The length of the films is 40 $\xi$.}
\end{figure}

Notice that the local minimum of $U_{max}(H)$ occurs at $H\simeq H_{c1}$ for all studied films (w=2-20 $\xi$) which is in contrast with dependence $I_c(H)$ for relatively wide films where position of dip is shifted to larger magnetic fields (see Fig. 2(b)) and approaches $H_s/2$ when $w \gg \xi$. This result is not surprising because in relatively wide films where $H_{c1} \lesssim H_s/2$ the vortices are washed out from the film by the transport current at $H_{c1}$ when $I\to I_c(H)$ (see discussion below Fig. 4) and dip may appear only at larger fields.
\begin{figure}[hbtp]
\includegraphics[width=0.45\textwidth]{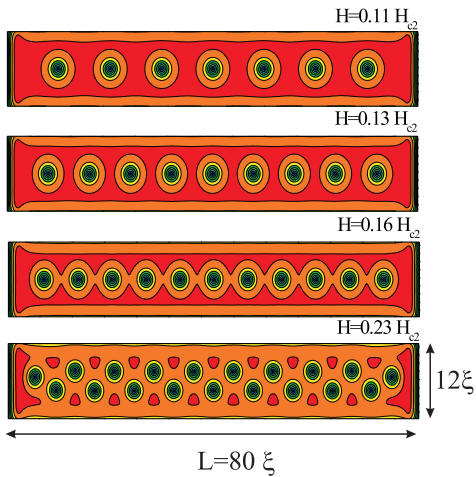}
\caption{Distribution of $|\Delta|$ in the film with $w=12 \xi$
and $L=80 \xi$ being in the ground state at different magnetic fields (I=0, compare it with Fig. 5 where $I\sim I_c(H)$).}
\end{figure}

\subsection{Comparison with the London model}

For relatively wide films the energy barrier $U=U_{en}$ at $H<H_0$ decays almost linearly with $H$ (see Fig. 14) which coincides with the predictions of the London model (see for example \cite{Steijic,Maksimova}). For relatively narrow films $U_{en}$ decays nonlinearly with $H$ which
reflects the contribution of magnetic field dependent vortex core energy $U_{core}(H)$ to $U_{en}$ \cite{Vodolazov3} (compare distribution of $|\Delta|$ in Fig. 12 at $H=0.1 H_{c2}$ and $H=0.7 H_{c2}$) while in the London model $U_{core}(H)=const\simeq 0.38 F_0$
\cite{Steijic}.

\subsection{Comparison with the experiments}

There are many works \cite{Parks,Patel,Wang,Cordoba,Herzog,Johansson,Parendo} where measured $R(H)$ has shape similar to curve $R^a$ in Fig. 1. There is simple criteria to distinguish, to which of these results the present here theory could be relevant - according to our calculations the first (or single) peak in $R(H)$ should occur at $H=H_{c1}$ which is shown in Fig. 8. The works \cite{Parks,Patel,Wang,Cordoba} relatively well fit to this condition but the better agreement with a theory is reached when one uses a little smaller value of the width. Two more experiments
\cite{Herzog,Johansson} also could be related to the present theory although the peak in R(H) occurs at considerably larger magnetic field
$\sim 2.3 H_{c1}$. It is interesting to note that in Refs. \cite{Herzog,Johansson} different materials were used (Sn and a:InO correspondingly)
but first peak in R(H) occurs almost at the same magnetic field for films with
comparable widths (compare Fig. 3(b) in Ref. \cite{Herzog} with Fig. 2 in Ref. \cite{Johansson}). Main difference between these experiments
is in the presence of short period oscillations in $R(H)$ observed in \cite{Johansson} and their period $\Delta H \sim 2 \Phi_0/Lw$ is close
to ours for film with $w \gtrsim 10 \xi$.

As we show in subsection II.B the locally smaller $T_c$ along the edges shifts the position of the dip in dependence $I_c(H)$ to the larger fields.
In Fig. 16 we demonstrate that $H_{c1}$ increases when the order parameter near the edges decreases due to lower value of $T_c$.
Effect is clearly stronger in relatively narrow film with $w=5 \xi$ where decrease of the 'superconducting' width by $\sim 2\xi$ (
length scale of proximity effect) have strong effect on $H_{c1}$. This result shows that smaller 'superconducting' width than the
real width of the film could be the reason for quantitative disagreement between the theory and the experiment.
\begin{figure}[hbtp]
\includegraphics[width=0.52\textwidth]{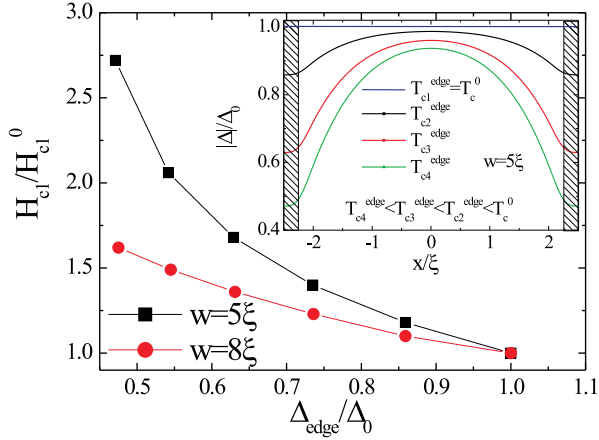}
\caption{Dependence of $H_{c1}$ for the films with nominal widths $w=5 \xi$ and $8\xi$ on the level of suppression of superconductivity near the edges due to locally smaller $T_c$. In the inset we present distribution of $|\Delta|$ across the film for different $T_c^{edge}$ at H=0 and
I=0. Shadowed areas mark the region where $T_c$ is locally suppressed in our model.}
\end{figure}

Another source of quantitative discrepancy between the theory and some experiments may come from no rectangular geometry.
In Ref. \cite{Parks} the superconducting film was placed on the surface of cylinder while in Ref. \cite{Patel} the cylindrical
nanowires were studied which raises a question about effective width of such a samples and correct value of $H_{c1}$.

Note that in some cases similar in shape dependence $R(H)$ cannot be explained by vortex assisted resistivity. For example
in Ref. \cite{Parendo} qualitatively similar dependence $R(H)$ was observed but position of the peak occurs at
$H \simeq 10^{-4} \Phi_0/w^2 \ll H_{c1}$ and we conclude that nonmonotonic $R(H)$ has a different origin.

\begin{figure}[hbtp]
\includegraphics[width=0.45\textwidth]{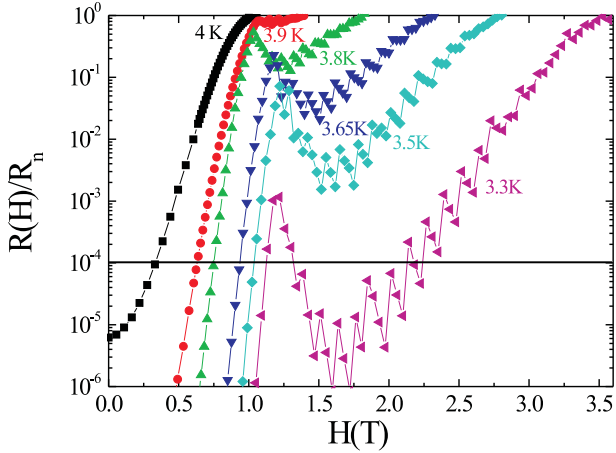}
\caption{Magnetoresistance of the superconducitng film with parameters of Ref. \cite{Cordoba} calculated with help of Eq. (4) and numerical results for $U_{max}(H)$ (part of which is present in Fig. 13). $w=2 \xi(T)$ at T=4 K and $w=4\xi(T)$ at T=3.3 K. Solid horizontal line shows the lowest measured resistance in Ref. \cite{Cordoba}.}
\end{figure}

In Fig. 17 we plot $R(H)$ which is calculated for
parameters of tungsten film ($\xi(0)$=6 nm, $\lambda(0)$=640 nm, w=50 nm, d=30 nm) from Ref. \cite{Cordoba} and where we assume Ginzburg-Landau temperature dependence for $\xi(T)=\xi(0)/(1-T/T_c)^{1/2}$ and $\lambda(T)=\lambda(0)/(1-T/T_c)^{1/2}$. We find the temperatures where the width of the film reaches $2\xi$, $2.5 \xi$, $3\xi$, $3.5 \xi$ and $4\xi$ and when insert in Eq. (4) numerically calculated $U_{max}(H)$ (parameter
$F_0/k_BT$ varies from 20 at T=4 K (w=2$\xi$) up to 94 at T=3.3 K (w=4 $\xi$)). Note that at calculations we did not use any fitting parameters and nevertheless find qualitative agreement with results of Ref. \cite{Cordoba} (compare Fig. 17 with Fig. 2(a) from Ref. \cite{Cordoba}).
In Ref. \cite{Cordoba} short period oscillations of $R(H)$ were not observed probably because of very large length of the sample
$L\simeq 4 \mu m \simeq 670 \xi(0)$. Quantitative agreement becomes better if one uses smaller width of the film in theoretical calculations
(it shifts local maximum of theoretical R(H) closer to experimental values and relative change of resistance
at given temperature becomes closer to experimental findings).

\section{Conclusion}

In framework of Ginzburg-Landau model it is shown that transport properties (critical current $I_c$ and resistance $R$ due to thermoactivated vortex hopping via energy barriers at $I \ll I_c$) of long narrow superconducting films with width of about several $\xi$ varies
nonmonotonically with external magnetic field. Due to appearance of the vortex row in the film critical current increases and resistance decreases at $H \gtrsim H_{c1}$ until the intervortex distance in the row becomes smaller than $w$. At larger magnetic fields $I_c$
decreases while R increases. Effect is most strong in films with width $w \simeq 3-8 \xi$. In wider films, already at fields when first vortex row appears in the superconductor the intervortex distance becomes less or comparable with $w$ and found effect is practically washed out due to vortex-vortex interaction.

Comparison with experiments demonstrates good qualitative agreement for position of the dip/peak in dependence $I_c(H)$ (R(H)) and in evolution of shape of $R(H)$ with temperature. Agreement becomes quantitative if one uses smaller width of the superconductor which looks reasonable
of one assumes uniform (along the film) degradation of the superconducting properties (lower value of $T_c$) near the edges. This degradation weekly influences the existence of the peak effect and only shifts the position of the peak to larger fields and affects its amplitude.

Contrary, the variations of the physical properties (width and/or $T_c$) along the film has a destructive impact on the peak effect if these variations are relatively large. In such a films the minimum of $I_c(H)$ (maximum of $R(H)$) occurs at different fields $H^*\sim \Phi_0/w^2$ (corresponding to different 'superconducting' widths in various parts of the film)
and it smears out one well pronounced dip/peak.

Finite length of the film produces additional short period oscillations both in $I_c(H)$ and $R(H)$ which are connected with discrete change in the number of the vortices. Amplitude of these oscillations decreases with increasing length of the film but it is still noticeable for films with length $40 \xi$. Period of these oscillations is in quantitative agreement with experimental findings of Ref. \cite{Johansson}.

All present results are found in framework of the GL model and are assumed to be quantitatively valid only close to $T_c$. But we do not expect large quantitative difference and for lower temperatures (at least for 'dirty' superconductors). For example, calculations of the energy barrier for phase slip event in 1D superconductor at arbitrary temperatures (using Usadel equation) \cite{Semenov} revealed small difference with result found in the GL model \cite{Langer} (if one uses proper temperature dependence for $\xi(T)$ and $\lambda(T)$ at low temperatures). Besides both the peak effect and negative magnetoresistance are most noticeable in relatively narrow films at fields $H \gtrsim H_{c1} \sim H_c$ and the order parameter at these magnetic fields is well suppressed below equilibrium value which justifies, in some respect, the usage of the GL model at lower temperatures.

\begin{acknowledgments}

Author thanks A.S. Mel'nikov and N.B. Kopnin for fruitful discussions. Author also thanks V. M. Vinokur for very helpful dispute about relation between the maximal energy barrier and probability for vortex passage across the superconducitng film. The work was supported by the Russian Foundation for Basic Research (project 12-02-00509) and by The Ministry of education and science of Russian Federation (project 8686).
\end{acknowledgments}


\end{document}